\documentclass[onecolumn,tightenlines,nofootinbib,preprintnumbers,
amsmath,amssymb]{revtex4}

\usepackage[usenames,dvipsnames]{color}
\usepackage{caption2}
\usepackage{dcolumn}
\usepackage{graphicx}
\usepackage{subfigure}
\usepackage{epstopdf}
\usepackage{bm}
\usepackage{lscape}
\renewcommand{\arraystretch}{0.9}
\usepackage{setspace}

\begin{document}
\begin{spacing}{1.5}

\title{Resonant contribution of the three-body decay process $\bar B_{s}  \rightarrow K^{+}K^{-} P$ in perturbation QCD}

\author{Gang L\"{u}$^{1}$\footnote{Email: ganglv66@sina.com}, Chang Chang Zhang $^{1}$\footnote{Email: 1219765284@qq.com}, Yan-Lin Zhao$^{1}$\footnote{Email:zyl163mail@163.com}, Li-Ying Zhang $^{1}$\footnote{Email:zhangly0324@163.com}}

\affiliation{\small $^{1}$College of Science, Henan University of Technology, Zhengzhou 450001, China\\
}

\begin{abstract}
We investigate the CP violation in the decay process  $\bar B_{s} \rightarrow \phi(\rho,\omega) P \rightarrow K^{+}K^{-}P$ by considering the interference effects of  $\phi\rightarrow K^{+}K^{-}$, $\rho\rightarrow K^{+}K^{-}$ and $\omega\rightarrow K^{+}K^{-}$ within the framework of  perturbative QCD method (P refers to $\pi$, K, $\eta$ and $\eta'$ pseudoscalar mesons, respectively). We analyse the mixings of $\phi-\rho^{0}$, $\phi-\omega$ and $\omega-\rho^{0}$ and provide the amplitudes of the quasi-two-body decay processes. The CP violation for $\bar B_{s} \rightarrow K^{+}K^{-} P$ decay process is obvious at the ranges of the three vector mesons interferences. Meanwhile, the localised CP violation can be found for comparing with the experiment results from three-body decay process at the LHC in the near future.

\end{abstract}

\maketitle

\section{Introduction}
\label{sec:sample1}
CP violation is a fascinating phenomenon in particle physics that has puzzled us for decades. The Standard Model (SM) of particle physics provides a framework for understanding CP violation, but there are still many unanswered questions \cite{Cabibbo:1963yz}.
One area of research focuses on the search for new sources of CP violation beyond the Cabibbo-Kobayashi-Maskawa (CKM) matrix. This involves studying rare decays and interactions between particles to look for deviations from the predictions of the Standard Model.
Another approach is to study CP violation in different types of particles, such as neutrinos or mesons.
Despite these efforts, much remains unknown about CP violation.

As early as 2012, LHCb Collaboration confirmed the existence of CP violation in some three-body decay studies of B mesons and found that the local phase space of $\bar B^{\pm} \rightarrow \pi^{+}\pi^{-}\pi^{\pm}$ decay channels had large direct CP violation, which was an interesting phenomenon at the time \cite{LHCb:2012kja, LHCb:2012uja}.
This phenomenon was later found to be explained by intermediate state resonances between different isospin mesons. As the $\bar B^{\pm} \rightarrow \pi^{+}\pi^{-}\pi^{\pm}$ decay process was studied using $\rho-\omega$ mixed resonance and found significant CP violation in the invariant mass m($\pi^{+}\pi^{-}$)=0.77GeV, which coincides with the position and degree of local CP violation \cite{Wang:2015ula}. There is no doubt that the three-body decay of heavy mesons is more complex than the two-body case, and one of the reasons is that they receive both resonant and non-resonant contributions during the decay process. The existing experimental results show that CP asymmetry in some local regions of phase space may be more obvious. Just like the LHCb observed large asymmetries in local regions in $B^{ \pm} \rightarrow K^{ \pm} \pi^{+} \pi^{-}$ and  $B^{ \pm} \rightarrow K^{ \pm} K^{+} K^{-}$.
Their invariant mass spectra of   $B^{ \pm} \rightarrow K^{ \pm} \pi^{+} \pi^{-}$ decays in the region  $0.08<m_{\pi^{+} \pi^{-}}^{2}<0.66 \mathrm{GeV}^{2} / c^{4} $ and $ m_{K^{ \pm} \pi^{\mp}}^{2}<   15 \mathrm{GeV}^{2} / c^{4} $, and   $B^{ \pm} \rightarrow K^{ \pm} K^{+} K^{-}$ decays in the region  $1.2<m_{K^{+} K^{-} \text {low }}^{2}<2.0 \mathrm{GeV}^{2} / c^{4} $ and  $m_{K^{+} K^{-} \text {high }}^{2}<15 \mathrm{GeV}^{2} / c^{4}$ \cite{LHCb:2013ptu}. These local apparent CP asymmetries are interesting. Currently, the phenomenon of CP asymmetry in the three-body decay process of $B_s$ mesons remains relatively unexplored, with limited research from both theoretical and experimental perspectives.

This paper aims to calculate the CP violation of $\bar B_{s} \rightarrow K^{+}K^{-} P$ decay process under the perturbative QCD method (PQCD). The reason is that the Sudakov factor in PQCD  effectively depresses the non-perturbative contribution and absorbs the non-perturbative part into the universal hadronic wave function \cite{Ali:2007ff}. Besides, this method is  self-consistent in the two-body non-leptonic decay process of B meson and has been proved to consisted  with the large CP violation found in experiment \cite{xiao2007xc}. Indeed, the corresponding two-body decay process of the B meson has been well-established and developed into  various of three-body decay process which we can treat three-body decay process with the method of quasi-two body decay process \cite{Hua:2020usv, Zou:2020fax}. In recent years, an increasing number of analysis about precious measurements of the branching ratio and CP violation in the three-body decay process have been carried out by BaBar \cite{BaBar:2014zli}, Belle II \cite{Bertacchi:2023jzv}, CLEO \cite{CLEO:2007vpk}, and LHCb \cite{Aaijprl2013}, which provides a great platform to test the standard Model (SM) and search the new physical signals. In this paper, we take the method of quasi-two-body decay process to calculate the CP violation of $\bar B_{s} \rightarrow K^{+}K^{-} P$ process under the mixing mechanism of $\phi\rightarrow K^{+}K^{-}$, $\rho^{0}\rightarrow K^{+}K^{-}$ and $\omega\rightarrow K^{+}K^{-}$.
 The reasons to explore the resonance effect among the three
particles arises from the adjacent masses of $\phi(1020)$, $\omega$(782) and $\rho^{0}$(770).
By incorporating information on $K^{+}K^{-}$ production and taking into account the constraints imposed by isospin symmetry, quark model and OZI rule, it becomes feasible to disentangle amplitudes with isospin
 $I=1$ and $ I=0$ components.
 The $\phi(1020)$ and $\omega$(782) match the isospin $I=0$ component.
The $I=1$ component  derives from $\rho^{0}(770)$.
 The ideal field of intermediate states is transformed into a computable physical field through the application of a unitary matrix in this paper. Additionally, we investigate localized CP violation within the hybrid resonance range to facilitate meaningful future comparisons with experimental results.

We present our work in six distinct parts.
The mechanism of three vector mesons mixing is introduced in section 2.
 In Section 3, we initially investigate CP violation arising from the involvement of the mixing mechanism in the decay process $\bar B_{s} \rightarrow \phi$ $(\rho^{0}, \omega)$ $ P  \rightarrow K^{+}K^{-} P$. Subsequently, we present a formalism for local CP violation. In Section 4, we introduce the amplitude formalism within the framework of perturbative QCD (PQCD) method, along with the fundamental functions and associated parameters. Additionally, we provide an evaluation of both the magnitude and integrated form of CP violation.
The analysis of data results can be found in Section 5. Finally, we engage in a comprehensive discussion and provide a concise summary of our findings.

\section{The mechanism of three vector mesons mixing}
The positive and negative electrons annihilate into photons and then they are polarized in a vacuum to form the mesons of $\phi (1020)$, $\rho^0(770)$ and $\omega(782)$, which can also decay into $K^{+} K^{-}$ pair. Meanwhile, the momentum can also be passed through the VMD model \cite{PMplb1981,Achasov2016}.
 Since the intermediate state particle is an un-physical state, we need convert it into a physical field from an isospin field through the matrix R \cite{Lu:2022rdi}. Then we can obtain the physical state of $\phi$, $\rho^{0}$ and $\omega$. What deserved to mentioned is that there is no $\phi- \rho^{0}- \omega$ mixing in the physical state and we neglect the contribution of the high-order term \cite{Lv2023epj}.
 The physical states $\phi- \rho^{0}- \omega$
 can be expressed as linear combinations of the isospin states
 $\phi_{I}- \rho^{0}_{I}- \omega_{I}$.
 The relationship can be represented by the following matrix:

\begin{equation}
	\left (
	\begin{array}{lllll}
		\rho^0\\[0.5cm]
		\omega\\[0.5cm]
		\phi
	\end{array}
	\right )
	=
	R(s)
	\left (
	\begin{array}{lll}
		\rho^0_I\\[0.5cm]
		\omega_I\\[0.5cm]
		\phi_I
	\end{array}
	\right )
	\label{L1}
\end{equation}
\noindent
where
\begin{equation}
	R  =
	\left (
	\begin{array}{lll}
		<\rho_{I}|\rho> & \hspace{2.cm} <\omega_{I}|\rho>  &\hspace{2.cm}<\phi_{I}|\rho>\\[0.5cm]
		<\rho_{I}|\omega> &  \hspace{2.cm}<\omega_{I}|\omega>&\hspace{2.cm}<\phi_{I}|\omega>\\[0.5cm]
		<\rho_{I}|\phi>&\hspace{2.cm} <\omega_{I}|\phi> & \hspace{2.cm} <\phi_{I}|\phi>
	\end{array}
	\right ).\\
	\label{L2}
\end{equation}
The change between the physical field and the isospin field in the intermediate state of the decay process is related by the matrices R.
The off-diagonal elements of R present the information of $\phi- \rho^{0}- \omega$ mixing.
Based on the isospin representation of $\phi_{I}$, $\rho_{I}$ and $\omega_{I}$, the isospin vector $|I,I_{3}>$ can be constructed,
where $I_3$ denotes the third component of isospin.
The variables i and j are employed to denote the physical state of the particle and the isospin basis vector, respectively. According to the orthogonal normalization relationship, we can derive:
$
\sum_{j}|j><j|=\sum_{j_{I}}\left|j_{I}><j_{I}\right|=I,
$ and $
<j\left|i>=<j_{I}\right| i_{I}>=\delta_{j i}
$.
We use the notation $F_{V_iV_j}$ to denote the mixing parameter, where $V_i$ and $V_j$ represent one of the three vector particles. Then,
the transformation matrix R can be converted as follows:
\begin{equation}
 R=\left(\begin{array}{ccc}
1 & -F_{\rho \omega}(s) & -F_{\rho \phi}(s) \\
F_{\rho \omega}(s) & 1 & -F_{\omega \phi}(s) \\
F_{\rho \phi}(s) & F_{\omega \phi}(s) & 1
\end{array}\right).
\end{equation}

From the translation of the two representations, the physical states can be written as
\begin{equation}
	\begin{split}
		\phi=F_{\rho\phi  }(s) \rho_{I}^{0}+F_{\omega \phi}(s) \omega_{I}+\phi_{I}, \\
		\omega=F_{ \rho\omega }(s) \rho_{I}^{0}+\omega_{I} -F_{\omega \phi}(s) \phi_{I}, \\
		\rho^{0}=\rho_{I}^{0}-F_{\rho\omega }(s) \omega_{I}-F_{\rho\phi }(s) \phi_{I}.
	\end{split}
\end{equation}
The relationship between the mixing parameters $\Pi_{V_{i} V_{j}}$ and $F_{V_{i}V_{j}}$ can be deduced from the subsequent equation:
\begin{equation}
	\begin{split}
		F_{\rho \omega}=\frac{\Pi_{\rho \omega}}{s_{\rho}-s_{\omega}},\\
		F_{\rho \phi}=\frac{\Pi_{\rho \phi}}{s_{\rho}-s_{\phi}},\\
		F_{\omega \phi}=\frac{\Pi_{\omega \phi}}{s_{\omega}-s_{\phi}}.
	\end{split}
\end{equation}
The relationship of $ F_{V_{i} V_{j}}$=$-F_{V_{j} V_{V_{i}}}$ can be found.
The inverse propagator of the vector meson, denoted as $s_V$ ($V = \phi, \rho$, or $\omega$), is defined such that $s_{V}=s-m_{V}^{2}+\mathrm{i} m_{V} \Gamma_{V}$.
The variables $m_V$ and $\Gamma_{V}$ represent the mass and decay rate of the vector mesons, respectively. Meanwhile, $\sqrt{s}$ denotes the invariant mass of the $K^{+}K^{-}$ pairs.

In this paper, the momentum dependence of the mixing parameters $\Pi_{V_{i} V_{j}}$ of $V_{i}V_{j}$ mixing is introduced to obtain the obvious s dependence.
The mixing parameter $\Pi_{\rho \omega }=-4470 \pm 250 \pm 160-i(5800 \pm 2000 \pm 1100)  \mathrm{MeV}^{2}$ is obtained near the $\rho$ meson is recently determined precisely by Wolfe and Maltnan \cite{CE2009,CE2011,Lu:2018fqe}. The mixing parameter $\Pi_{\omega \phi}=19000+i(2500 \pm 300) \mathrm{MeV}^{2}$ is obtained near the $\phi$ muon. And the mixing parameter $\Pi_{\phi\rho}=720 \pm 180 -i(870 \pm 320) \mathrm{MeV}^{2}$ is obtained near the $\phi$ meson \cite{ MN2000}. Then we define
\begin{eqnarray}
\widetilde{\Pi}_{\rho\omega}=\frac{s_{\rho}\Pi_{\rho\omega}}{s_{\rho}-s_{\omega}},
~~\widetilde{\Pi}_{\rho\phi}=\frac{s_{\rho}\Pi_{\rho\phi}}{s_{\rho}-s_{\phi}},
~~\widetilde{\Pi}_{\phi\omega}=\frac{s_{\phi}\Pi_{\phi\omega}}{s_{\phi}-s_{\omega}}.
\end{eqnarray}

\section{CP violation in $\bar B_{s} \rightarrow \phi$ ($\rho^{0}$, $\omega$) $ P  \rightarrow K^{+}K^{-} P$ decay process }
\subsection{The resonance effect from $V\rightarrow K^{+}K^{-}$ }

\label{sec:spectra}
\begin{figure}[h]
	\centering
	\includegraphics[height=8cm,width=12cm]{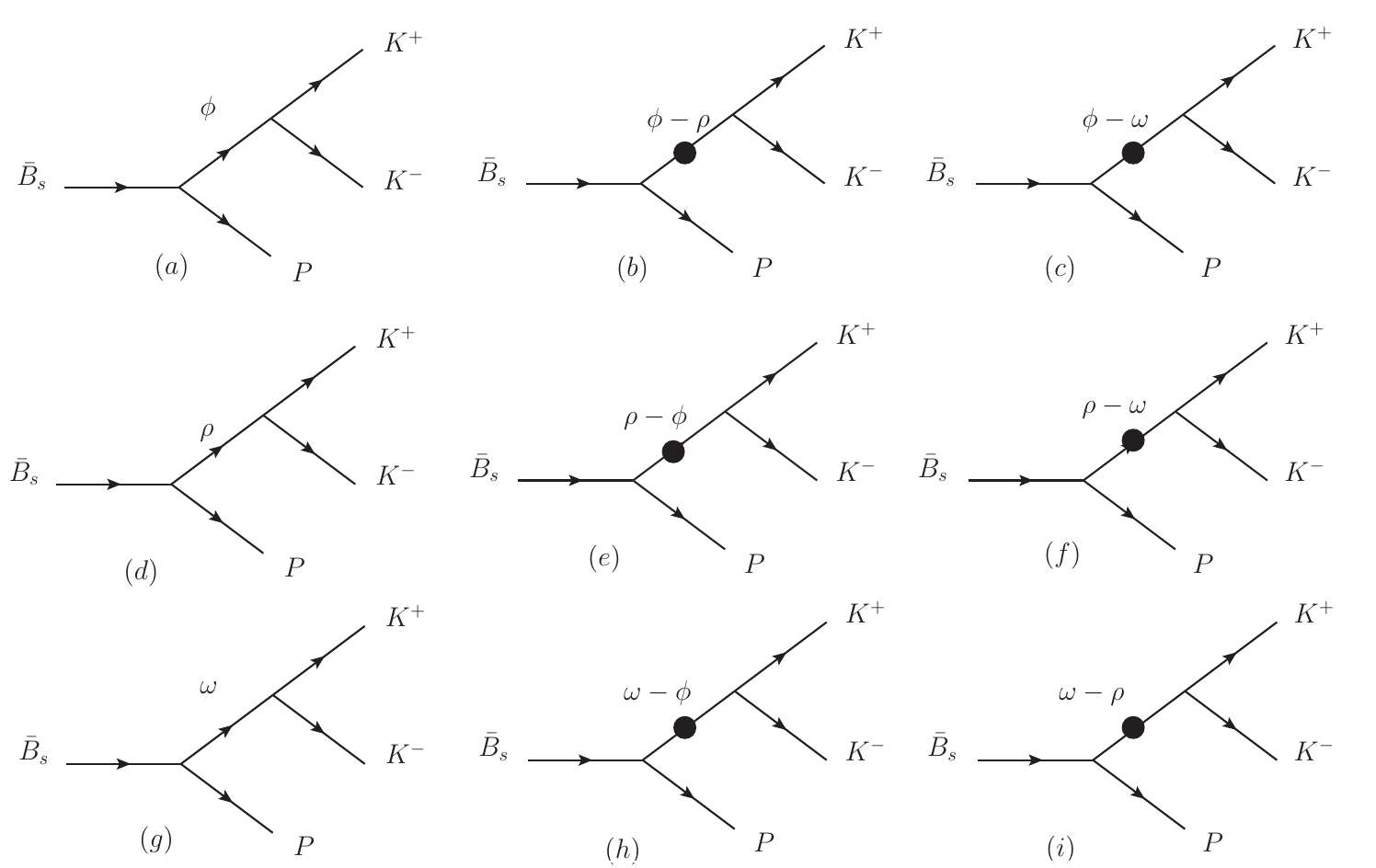}
	\caption{ The decay diagrams of $\bar B_{s} \rightarrow \phi $ $(\rho^{0}$, $\omega)$ $P \rightarrow K^{+}K^{-} $ $P$ process.}
	\label{fig1}
\end{figure}

We present decay diagrams (a)-(i) of the $\bar B_{s} \rightarrow \phi$ $(\rho^{0}$, $\omega)$ $P \rightarrow K^{+} K^{-} P$ process in Fig.\ref{fig1}, aiming to provide a more comprehensive understanding of the mixing mechanism.

In the above decay diagrams, the decay processes depicted in (a), (d), and (g) represent direct decay modes, where $K^{+} K^{-}$ are produced through $\phi$, $\rho^0$, and $\omega$ respectively. The quasi-two-body approach employed in this study is evident from the aforementioned diagrams.
Compared to the direct decay processes depicted in diagrams (a), (d), and (g) of Fig.\ref{fig1}, the $K^{+} K^{-}$ pair can also be generated through a distinct mixing mechanism. The black dots in the figure represent the resonance effect between these two mesons, denoted by the mixing parameter $\Pi_{V_{i} V_{j}}$. Although the contribution from this mixing mechanism is relatively small compared to other diagrams in Fig.\ref{fig1}, it must  be taken into consideration.

The amplitude of the $\bar{B}_{s} \rightarrow \phi $ ($\rho^{0}, \omega$) $ P  \rightarrow K^{+}K^{-} P $ decay channel can be characterized in the following manner:
\begin{equation}
	A=\left \langle K^{+}K^{-}P\left | H^{T} \right | \bar{B}_{s} \right \rangle +\left \langle K^{+}K^{-}P\left | H^{P} \right | \bar{B}_{s} \right \rangle,
\end{equation}
The quantities $\left \langle K^{+}K^{-}P\left | H^{P} \right | \bar{B}_{s} \right \rangle $ and $\left \langle K^{+}K^{-}P\left | H^{T} \right | \bar{B}_{s} \right \rangle $ represent the amplitudes associated with penguin-level and tree-level contributions, respectively.
The propagator of the intermediate vector meson can be transformed from the diagonal matrix to the physical state after applying the R matrix transformation. Neglecting higher order terms, the amplitudes
can be as demonstrated below:
\begin{eqnarray}
\begin{split}
\left\langle K^{+} K^{-} P\left|H^{T}\right| \bar B_{s}\right\rangle=
&
\frac{g_{\phi}}{s_{\phi}}t_{\phi}
+\frac{g_{\rho}}{s_{\rho}s_{\phi}}\widetilde{\Pi}_{\rho\phi}t_{\phi}
+\frac{g_{\omega}}{s_{\omega}s_{\phi}}\widetilde{\Pi}_{\omega\phi}t_{\phi}
+\frac{g_{\rho}}{s_{\rho}}t_{\rho}
+\frac{g_{\phi}}{s_{\phi}s_{\rho}}\widetilde{\Pi}_{\phi\rho}t_{\rho}
\\
&
+\frac{g_{\omega}}{s_{\omega}s_{\rho}}\widetilde{\Pi}_{\omega\rho}t_{\rho}
+\frac{g_{\omega}}{s_{\omega}}t_{\omega}
+\frac{g_{\phi}}{s_{\phi}s_{\omega}}\widetilde{\Pi}_{\phi\omega}t_{\omega}
+\frac{g_{\rho}}{s_{\rho}s_{\omega}}\widetilde{\Pi}_{\rho\omega}t_{\omega},
\label{Htr}
\end{split}
\end{eqnarray}
\begin{eqnarray}
\begin{split}
\left\langle K^{+} K^{-} P\left|H^{P}\right| \bar B_{s}\right\rangle=
&
\frac{g_{\phi}}{s_{\phi}}p_{\phi}
+\frac{g_{\rho}}{s_{\rho}s_{\phi}}\widetilde{\Pi}_{\rho\phi}p_{\phi}
+\frac{g_{\omega}}{s_{\omega}s_{\phi}}\widetilde{\Pi}_{\omega\phi}p_{\phi}
+\frac{g_{\rho}}{s_{\rho}}p_{\rho}
+\frac{g_{\phi}}{s_{\phi}s_{\rho}}\widetilde{\Pi}_{\phi\rho}p_{\rho}
\\
&
+\frac{g_{\omega}}{s_{\omega s_{\rho}}}\widetilde{\Pi}_{\omega\rho}p_{\rho}
+\frac{g_{\omega}}{s_{\omega}}p_{\omega}
+\frac{g_{\phi}}{s_{\phi}s_{\omega}}\widetilde{\Pi}_{\phi\omega}p_{\omega}
+\frac{g_{\rho}}{s_{\rho}s_{\omega}}\widetilde{\Pi}_{\rho\omega}p_{\omega},
\label{Hpe}
\end{split}
\end{eqnarray}
where the tree-level (penguin-level) amplitudes $t_{\rho}\left(p_{\rho}\right)$, $t_{\omega}\left(p_{\omega}\right)$, and $t_{\phi}\left(p_{\phi}\right)$ correspond to the decay processes $\bar B_s \rightarrow \rho^0 P$, $\bar B_s \rightarrow \omega P$ and $\bar B_s \rightarrow \phi P$, respectively. Here, $s_V$ represents the inverse propagator of the vector meson V  \cite{ Chen:1999nxa, Wolfe:2009ts, Wolfe:2010gf}.
Moreover, $g_{V}$ represents the coupling constant derived from the decay process of $ V \rightarrow K^{+} K^{-}$ and can be expressed as $\sqrt{2}g_{{\rho}k^{+} k^{-}}=\sqrt{2}g_{\omega k^{+} k^{-}}=-g_{\phi k^{+} k^{-}}=4.54$ \cite{Bruch:2004py}.

 The differential parameter for CP asymmetry can be expressed as follows:
\begin{equation}
	\label{cp31}
	A_{CP}=\frac{\left| A \right|^2-\left| \overline{A} \right|^2}{\left| A \right|^2+\left| \overline{A} \right|^2}.
\end{equation}

\subsection{The localised CP violation  of $A^{\Omega}_{CP}$}

 In this paper, we perform the integral calculation of A$_{CP}$ to facilitate future experimental comparisons. For the decay process $\bar B_{s} \rightarrow \phi P$, the amplitude is given by $M_{\bar B _{s}\rightarrow \phi P}^{\lambda}=\alpha p_{\bar B} \cdot \epsilon^{*}(\lambda)$, where $p_{\bar B_{s}}$ represents the momenta of the $\bar B_{s}$ meson, $\epsilon$ denotes the polarization vector of $\phi$ and $\lambda$ corresponds to its polarization. The parameter $\alpha$ remains independent of $\lambda$. Similarly, in the decay process $\phi \rightarrow K^{+}K^{-}$, we can express $M_{\phi \rightarrow K^{-} K^{+}}^{\lambda}=g_{\phi}\epsilon(\lambda)\left(p_1-p_2\right)$, where $p_1$ and $p_2$ denote the momenta of the produced $K^{+}$ and $K^{-}$ particles from $\phi$, respectively. Here, the parameter $g_\phi$ represents an effective coupling constant for $\phi \rightarrow K^{+}K^{-}$. Regarding the dynamics of meson decay, it is observed that the polarization vector of a vector meson satisfies $\sum_{\lambda=0,\pm 1}\epsilon^\lambda_\mu(p)(\epsilon^\lambda_\nu(p))^*=-(g_{\mu\nu}-p_\mu p_\nu/m_V^2)$. As a result, we obtain the total amplitude for the decay process $\bar B_s \rightarrow \phi P\rightarrow K^{+}K^-P$
   \cite{Guo:2000uc,Zhang:2013oqa,Wang:2015ula}:
\begin{equation}
	\begin{aligned}
		A &=\alpha p_{\bar B_{s}}^{\mu} \frac{\sum_{\lambda} \epsilon_{\mu}^{*}(\lambda) \epsilon_{\nu}(\lambda)}{s_{\phi}} g_{\phi kk}\left(p_{1}-p_{2}\right)^{\nu} \\
		&=\frac{g_{\phi kk} \alpha}{s_{\phi}} \cdot p_{\bar B_{s}}^{\mu}\left[g_{\mu \nu}-\frac{\left(p_{1}+p_{2}\right)_{\mu}\left(p_{1}+p_{2}\right)_{\nu}}{s}\right]\left(p_{1}-p_{2}\right)^{\nu} \\
		&=\frac{g_{\phi kk}}{s_{\phi}} \cdot \frac{M_{\bar B_{s}\rightarrow \phi \pi^{0}}^{\lambda}}{p_{\bar B_{s}} \cdot \epsilon^{*}} \cdot\left(\Sigma-s^{\prime}\right) \\
		&=\left(\Sigma-s^{\prime}\right) \cdot \mathcal{A}.
	\end{aligned}
\end{equation}
 The high ($\sqrt{s^\prime}$) and low $\sqrt{s}$ ranges are defined for calculating the invariant mass of $K^{-} K^{+}$. By setting a fixed value for $s$, we can determine an appropriate value for $s^\prime$ that fulfills the equation $\Sigma=\frac{1}{2}\left(s_{\max }^\prime+s_{\min }^\prime\right)$, where  ${ s}_{ \max  }^{ \prime  }( { s}_{ \min  }^{\prime}) $ denotes respectively the maximum (minimum) value.

 Utilizing the principles of three-body kinematics, we can deduce the local CP asymmetry for the decay $\bar{B}_{s} \rightarrow K^{+}K^{-} P $ within a specific range of invariant mass:
\begin{equation}
	A_{C P}^{\Omega}=\frac{\int_{s_{1}}^{s_{2}} \mathrm{~d} s \int_{s_{1}^{\prime}}^{s_{2}^{\prime}} \mathrm{d} s^{\prime}\left(\Sigma-s^{\prime}\right)^{2}\left(|\mathcal{A}|^{2}-|\overline{\mathcal{A}}|^{2}\right)}{\int_{s_{1}}^{s_{2}} \mathrm{~d} s \int_{s_{1}^{\prime}}^{s_{2}^{\prime}} \mathrm{d} s^{\prime}\left(\Sigma-s^{\prime}\right)^{2}\left(|\mathcal{A}|^{2}+|\overline{\mathcal{A}}|^{2}\right)}.
\end{equation}
Our calculation takes into account the dependence of $\Sigma=\frac{1}{2}\left(s_{\max }^{\prime}+s_{\min }^{\prime}\right)$ on $s^{\prime}$. Assuming that $s_{\max }^{\prime}>s^{\prime}>s_{\min }^{\prime}$ represents an integral interval of high invariant mass for the $K^{-} K^{+}$ meson pair, and $\int_{s_{1}^{\prime}}^{s_{2}^{\prime}} \mathrm{d}s^\prime(\Sigma-s')^{2}$ represents a factor dependent on $s'$. The correlation between $\Sigma$ and $s'$ can be easily determined through kinematic analysis, as $s'$ only varies on a small scale. Therefore, we can consider $\Sigma$ as a constant. This allows us to cancel out the term $\int_{s_1^\prime}^{ s_2^\prime }\mathrm{d}s^\prime (\Sigma-s')^{2}$ in both the numerator and denominator, resulting in $A_{C P}^{\Omega}$ no longer depending on the high invariant mass of positive and negative particles.

\section{\label{sec:cpv1}The amplitudes of quasi-two-body decay processes within the framework of perturbative QCD (PQCD)}

\subsection{Formulation of calculations}

The three-body decay process is accompanied by intricate and multifaceted dynamical mechanisms. The perturbative QCD (PQCD) method is known for its efficacy in handling perturbation corrections, which has been successfully applied to two-body non-light decay processes and holds promise for quasi-two-body decay processes as well.
In the framework of PQCD, within the rest frame of a heavy B meson, the decay process involves the production of two light mesons with significantly large momenta that exhibit rapid motion. The dominance of hard interactions in this decay amplitude arises due to insufficient time for exchanging soft gluons with the final-state mesons. Given the high velocity of these final-state mesons, a hard gluon imparts momentum to the light spectator quark within the B meson, resulting in the formation of a rapidly moving final-state meson. Consequently, this hard interaction is described by six quark operators. The nonperturbative dynamics are encapsulated within the meson wave function, which can be extracted through experimental measurements. On the other hand, employing perturbation theory allows for computation of this aforementioned hard contribution. Quasi-two-body decay can be computed by defining the intermediate state of decay.

By employing the quasi-two-body decay method, the total amplitude of  $\bar B_{s} \rightarrow \phi $ ($\rho^{0}$, $\omega$) $\pi^{0}$ $\rightarrow K^{+}K^{-} \pi^{0} $ is composed of two components: $\bar B_{s} \rightarrow \phi $ ($\rho^{0}$, $\omega$) $\pi^{0} $ and $\phi $ ($\rho^{0}$, $\omega) \rightarrow K^{+}K^{-} $. In this study, we illustrate the methodology of quasi-two-body decay process using the example of $\bar B_{s}\rightarrow\phi\pi^0\rightarrow K^+K^-\pi^0$, based on the matrix elements involving  $V_{tb}$, $V_{ts}^{*}$ and  $V_{ub}$,$V_{ub}^{*}$.

\begin{equation}
\begin{aligned}
\sqrt{2}A\left(\bar{B}_{s} \rightarrow \pi^{0} \phi\left(\phi\rightarrow K^{+}K^{-}\right)\right)=&\frac{\left.G_{F} p_{\bar{B}_{s}} \cdot \sum_{\lambda=0, \pm 1} \epsilon(\lambda) g_{\phi} \epsilon^{*}(\lambda) \cdot\left(p_{k^+}-p_{k^-}\right)\right.}{\sqrt{2} s_{\phi}}\\
&\times \bigg\{V_{u b} V_{u s}^{*}\left[f_{\pi} F_{\bar B_{s} \rightarrow \phi}^{L L}(a_{2})+M_{\bar B_{s} \rightarrow \phi}^{L L}(C_{2})\right]\bigg.\\
&\left.-V_{t b} V_{t s}^{*}\left[f_{\pi} F_{\bar B_{s} \rightarrow \phi}^{L L}\left(\frac{3}{2} a_{9}-\frac{3}{2} a_{7}\right)+M_{\bar B_{s} \rightarrow \phi}^{L L}\left(\frac{3}{2} C_{8}+\frac{3}{2} C_{10}\right)\right]\right\},
\end{aligned}
\end{equation}
where $P_{\bar B_s}$, $p_{k^{+}}$ and $p_{k^{-}}$ are the momentum of $\bar B_s$, $K^{+}$ and $K^{-}$, respectively. $C_i$ ($a_i$)  is Wilson coefficient (associated Wilson coefficient), $\epsilon$ is the polarization of vector meson . $G_F$ is the Fermi constant. $f_{\pi}$ refers to the decay constants of $\pi$ \cite{Li:2006jv}.
Besides $F_{\bar B_{s} \rightarrow \phi}^{L L}$ and $M_{\bar B_{s} \rightarrow \phi}^{L L}$ represent emission graphs that are factorable and non-factorable. $F_{a n n}^{L L}$ and $M_{a n n}^{L L}$ represent annihilation graphs that are factorable and non-factorable. $LL$, $LR$, and $SP$ correspond to three flow structures \cite{Ali:2007ff}.

The additional representations of the three-body decay amplitudes that necessitate consideration for calculating CP violation through the mixed mechanism in this paper are as follows:

\begin{equation}
\begin{aligned}
2 A\left(\bar{B}_{s}^{0} \rightarrow\right.\left.\rho^{0}\left(\rho^{0} \rightarrow K^{+} K^{-}\right) \pi^{0}\right) =& \frac{G_{F} p_{\bar{B}_{s}^{0}} \cdot  \sum_{\lambda = 0, \pm 1}\epsilon(\lambda) g_{\rho} \epsilon^{*}(\lambda) \cdot\left(p_{k^{+}}-p_{k^{-}}\right)}{ \sqrt{2} s_{\rho}} \\
&\times \left\{ V_{u b} V_{u s}^{*}\left[f_{B_{s}} F_{a n n}^{L L}(a_{2})+M_{a n n}^{L L}(C_{2})+f_{B_{s}} F_{a n n }^{L L '}(a_{2})+M_{a n n}^{L L'}(C_{2})\right]     \right.  \\
&- V_{t b} V_{t s}^{*}\left[ f_{B_{s}} F_{a n n}^{L L}\left(a_{3}+a_{9}\right)\right.\left.-f_{B_{s}} F_{a n n}^{L R}\left(a_{5}+a_{7}\right)+M_{a n n}^{L L}\left(C_{4}+C_{10}\right)\right.\\
&-M_{a n n}^{S P}\left(C_{6}+C_{8}\right)+\left[\pi^{+} \leftrightarrow \rho^{-}\right] +f_{B_{s}} F_{a n n}^{L L'}\left(a_{3}+a_{9}\right)-f_{B_{s}} F_{a n n}^{L R'}\left(a_{5}+a_{7}\right) \\
&\left.\left.+M_{a n n}^{L L'}\left(C_{4}+C_{10}\right)-M_{a n n}^{S P'}\left(C_{6}+C_{8}\right) +\left[\rho^{+} \leftrightarrow \pi^{-}\right] \right] \right\}.
\end{aligned}
\end{equation}

\begin{equation}
\begin{aligned}
2A\left(\bar{B}_{s}^{0} \rightarrow \pi^{0}\omega\left(\omega \rightarrow K^{+} K^{-}\right) \right)=&\frac{G_{F} p_{\bar{B}_{s}^{0}} \cdot \sum_{\lambda=0, \pm 1} \epsilon(\lambda) g_{\omega} \epsilon^{*}(\lambda) \cdot\left(p_{k^+}-p_{k^-}\right)}{ \sqrt{2} s_{\omega}}  \\
\times & \left\{V_{u b} V_{u s}^{*} M_{a n n}^{L L}\left(c_{2}\right)-V_{t b} V_{t s}^{*}\left[M_{a n n}^{L L}\left(\frac{3}{2} c_{10}\right)-M_{a n n}^{S P}\left(\frac{3}{2} c_{8}\right)+\left[\pi^{0} \leftrightarrow \omega\right]\right]\right\}.
\end{aligned}
\end{equation}

\begin{equation}
\begin{aligned}
A\left(\bar{B}_{s}^{0} \rightarrow K^{0}\phi\left(\phi \rightarrow K^{+} K^{-}\right) \right)=&\frac{G_{F} p_{\bar{B}_{s}^{0}} \cdot \sum_{\lambda=0, \pm 1} \epsilon(\lambda) g_{\phi} \epsilon^{*}(\lambda) \cdot\left(p_{k^+}-p_{k^-}\right)}{ \sqrt{2} s_{\phi}}  \\
&\times \left\{- V_{t b} V_{t d}^{*}\left[f_{\phi} F_{B_{s} \rightarrow K}^{L L}\left(a_{3}+a_{5}-\frac{1}{2} a_{7}-\frac{1}{2} a_{9}\right)+f_{K} F_{B_{s} \rightarrow \phi}^{L L}\left(a_{4}-\frac{1}{2} a_{10}\right) \right.  \right.\\
&\left. \left.-f_{K} F_{B_{s} \rightarrow \phi}^{S P}\left(a_{6}-\frac{1}{2} a_{8}\right)+M_{B_{s} \rightarrow K}^{L L}\left(C_{4}-\frac{1}{2} C_{10}\right)+M_{B_{s} \rightarrow \phi}^{L L}\left(C_{3}-\frac{1}{2} C_{9}\right) \right.\right.\\
&\left.\left. -M_{B_{s} \rightarrow K}^{S P}\left(C_{6}-\frac{1}{2} C_{8}\right)-M_{B_{s} \rightarrow \phi}^{L R}\left(C_{5}-\frac{1}{2} C_{7}\right)+f_{B_{s}} F_{a n n}^{L L}\left(a_{4}-\frac{1}{2} a_{10}\right) \right.\right.\\
&\left.\left. -f_{B_{s}} F_{a n n}^{S P}\left(a_{6}-\frac{1}{2} a_{8}\right)+M_{a n n}^{L L}\left(C_{3}-\frac{1}{2} C_{9}\right)-M_{a n n}^{L R}\left(C_{5}-\frac{1}{2} C_{7}\right) \right]\right\}.
\end{aligned}
\end{equation}

\begin{equation}
\begin{aligned}
\sqrt{2}A\left(\bar{B}_{s}^{0} \rightarrow K^{0}\rho\left(\rho \rightarrow K^{+} K^{-}\right) \right)=&\frac{G_{F} p_{\bar{B}_{s}^{0}} \cdot \sum_{\lambda=0, \pm 1} \epsilon(\lambda) g_{\phi} \epsilon^{*}(\lambda) \cdot\left(p_{k^+}-p_{k^-}\right)}{ \sqrt{2} s_{\rho}}\\
& \times \left\{  V_{u b} V_{u d}^{*}\left[f_{\rho} F_{B_{s} \rightarrow K}^{L L}\left(a_{2}\right)+M_{B_{s} \rightarrow K}^{L L}\left(C_{2}\right)\right]- V_{t b} V_{t d}^{*}\left[ M_{B_{s} \rightarrow K}^{L R}\left(-C_{5}+\frac{1}{2} C_{7}\right)\right.\right.\\
& \left.+f_{\rho} F_{B_{s} \rightarrow K}^{L L}\left(-a_{4}+\frac{3}{2} a_{7}+\frac{1}{2} a_{10}+\frac{3}{2} a_{9}\right)-M_{B_{s} \rightarrow K}^{S P}\left(\frac{3}{2} C_{8}\right) \right.\\
&\left.\left. +M_{B_{s} \rightarrow K}^{L L}\left(-C_{3}+\frac{1}{2} C_{9}+\frac{3}{2} C_{10}\right)+f_{B_{s}} F_{a n n}^{L L}\left(-a_{4}+\frac{1}{2} a_{10}\right) \right.\right.\\
& \left.\left.+f_{B_{s}} F_{a n n}^{S P}\left(-a_{6}+\frac{1}{2} a_{8}\right)+M_{a n n}^{L L}\left(-C_{3}+\frac{1}{2} C_{9}\right)+M_{a n n}^{L R}\left(-C_{5}+\frac{1}{2} C_{7}\right)\right]\right\}.
\end{aligned}
\end{equation}

\begin{equation}
\begin{aligned}
\sqrt{2}A\left(\bar{B}_{s}^{0} \rightarrow K^{0}\omega\left(\omega \rightarrow K^{+} K^{-}\right) \right)=&\frac{G_{F} p_{\bar{B}_{s}^{0}} \cdot \sum_{\lambda=0, \pm 1} \epsilon(\lambda) g_{\omega} \epsilon^{*}(\lambda) \cdot\left(p_{k^+}-p_{k^-}\right)}{ \sqrt{2} s_{\omega}} \\
  &\times  \left\{V_{u b} V_{u d}^{*}  \left[ f_{\omega} F_{B_{s} \rightarrow K}^{L L}\left(a_{2}\right)+M_{B_{s} \rightarrow K}^{L L}\left(C_{2}\right)\right] - V_{t b} V_{t d}^{*}\left[M_{B_{s} \rightarrow K}^{L R}\left(C_{5}-\frac{1}{2} C_{7}\right) \right.\right.\\
&\left.+f_{\omega} F_{B_{s} \rightarrow K}^{L L}\left(2 a_{3}+a_{4}+2 a_{5}+\frac{1}{2} a_{7}+\frac{1}{2} a_{9}-\frac{1}{2} a_{10}\right)\right.\\
& +M_{B_{s} \rightarrow K}^{L L}\left(C_{3}+2 C_{4}-\frac{1}{2} C_{9}+\frac{1}{2} C_{10}\right)+M_{a n n}^{L L}\left(C_{3}-\frac{1}{2} C_{9}\right)\\
&\left. -M_{B_{s} \rightarrow K}^{S P}\left(2 C_{6}+\frac{1}{2} C_{8}\right)+f_{B_{s}} F_{a n n}^{L L}\left(a_{4}-\frac{1}{2} a_{10}\right) \right.\\
&\left.\left.+f_{B_{s}} F_{a n n}^{S P}\left(a_{6}-\frac{1}{2} a_{8}\right)+M_{a n n}^{L R}\left(C_{5}-\frac{1}{2} C_{7}\right) \right]\right\}.
\end{aligned}
\end{equation}

\begin{equation}
\begin{aligned}
A\left(\bar{B}_{s}^{0} \rightarrow \eta\phi\left(\phi \rightarrow K^{+} K^{-}\right) \right)=&\frac{G_{F} p_{\bar{B}_{s}^{0}} \cdot \sum_{\lambda=0, \pm 1} \epsilon(\lambda) g_{\phi} \epsilon^{*}(\lambda) \cdot\left(p_{k^+}-p_{k^-}\right)}{ \sqrt{2} s_{\phi}} \\
&\times   \left\{ \frac{\cos \theta}{\sqrt{2}}\bigg\{ V_{u b} V_{u s}^{*}\left[f_{n} F_{B_{s} \rightarrow \phi}^{L L}\left(a_{2}\right)+M_{B_{s} \rightarrow \phi}^{L L}\left(C_{2}\right)\right] \bigg.\right.\\
&- V_{t b} V_{t s}^{*}\left[f_{n} F_{B_{s} \rightarrow \phi}^{L L}\left(2 a_{3}-2 a_{5}-\frac{1}{2} a_{7}+\frac{1}{2} a_{9}\right)\right. \\
& \left.\left.+M_{B_{s} \rightarrow \phi}^{L L}\left(2 C_{4}+\frac{1}{2} C_{10}\right)+M_{B_{s} \rightarrow \phi}^{S P}\left(2 C_{6}+\frac{1}{2} C_{8}\right)\right]\right\}\\
&-\sin \theta\left\{- V_{t b} V_{t s}^{*}\left[f_{s} F_{B_{s} \rightarrow \phi}^{L L'}\left(a_{3}+a_{4}-a_{5}+\frac{1}{2} a_{7}-\frac{1}{2} a_{9}-\frac{1}{2} a_{10}\right)\right.\right. \\
& +M_{B_{s} \rightarrow \phi}^{S P'}\left(C_{6}-\frac{1}{2} C_{8}\right)+f_{B_{s}} F_{a n n}^{L L'}\left(a_{3}+a_{4}-a_{5}+\frac{1}{2} a_{7}-\frac{1}{2} a_{9}-\frac{1}{2} a_{10}\right) \\
& +M_{a n n}^{L L'}\left(C_{3}+C_{4}-\frac{1}{2} C_{9}-\frac{1}{2} C_{10}\right)-f_{B_{s}} F_{a n n}^{S P'}\left(a_{6}-\frac{1}{2} a_{8}\right) \\
& \left.\left.\left.-M_{a n n}^{L R'}\left(C_{5}-\frac{1}{2} C_{7}\right)-M_{a n n}^{S P'}\left(C_{6}-\frac{1}{2} C_{8}\right)\right]+\left[\eta_{s} \leftrightarrow \phi\right]\right\}\right\} .\\
\end{aligned}
\end{equation}

\begin{equation}
\begin{aligned}
A\left(\bar{B}_{s}^{0} \rightarrow \eta \rho^{0}\left( \rho^{0} \rightarrow K^{+} K^{-}\right) \right)=&\frac{G_{F} p_{\bar{B}_{s}^{0}} \cdot \sum_{\lambda=0, \pm 1} \epsilon(\lambda) g_{ \rho} \epsilon^{*}(\lambda) \cdot\left(p_{k^+}-p_{k^-}\right)}{ \sqrt{2} s_{ \rho}} \\
&\times   \left\{\frac{\cos \theta}{2}\left\{-V_{t b} V_{t s}^{*}\left[f_{B_{s}} F_{a n n}^{L L}\left(\frac{3}{2} a_{9}-\frac{3}{2} a_{7}\right)+M_{a n n}^{L L}\left(\frac{3}{2} C_{10}\right)-M_{a n n}^{S P}\left(\frac{3}{2} C_{8}\right)\right]\right.\right. \\
& \left.+ V_{u b} V_{u s}^{*}\bigg[f_{B_{s}} F_{a n n}^{L L}\left(a_{2}\right)+M_{a n n}^{L L}\left(C_{2}\right)\bigg]+\left[\rho^{0} \leftrightarrow \eta_{n}\right]\right\}\\
&-\frac{\sin \theta}{\sqrt{2}}\left\{ V_{u b} V_{u s}^{*}\bigg[f_{\rho} F_{B_{s} \rightarrow \eta_{s}}^{L L'}\left(a_{2}\right)+M_{B_{s} \rightarrow \eta_{s}}^{L L'}\left(C_{2}\right)\bigg]\right. \\
&\left. \left.- V_{t b} V_{t s}^{*}\left[f_{\rho} F_{B_{s} \rightarrow \eta_{s}}^{L L'}\left(\frac{3}{2} a_{7}+\frac{3}{2} a_{9}\right)+M_{B_{s} \rightarrow \eta_{s}}^{L L'}\left(\frac{3}{2} C_{10}\right)-M_{B_{s} \rightarrow \eta_{s}}^{S P'}\left(\frac{3}{2} C_{8}\right)\right]\right\} \right\}.
\end{aligned}
\end{equation}

\begin{equation}
\begin{aligned}
A\left(\bar{B}_{s}^{0} \rightarrow \eta \omega\left( \omega \rightarrow K^{+} K^{-}\right) \right)=&\frac{G_{F} p_{\bar{B}_{s}^{0}} \cdot \sum_{\lambda=0, \pm 1} \epsilon(\lambda) g_{ \omega} \epsilon^{*}(\lambda) \cdot\left(p_{k^+}-p_{k^-}\right)}{ \sqrt{2} s_{ \omega}} \\
&\times   \left\{\frac{\cos \theta}{2}\left\{  V_{u b} V_{u s}^{*}\bigg[f_{B_{s}} F_{a n n}^{L L}\left(a_{2}\right)+M_{a n n}^{L L}\left(C_{2}\right)\bigg]\right.\right. \\
& - V_{t b} V_{t s}^{*}\left[M_{a n n}^{L L}\left(2 C_{4}+\frac{1}{2} C_{10}\right)-M_{a n n}^{S P}\left(2 C_{6}+\frac{1}{2} C_{8}\right)\right. \\
+ & \left.\left.f_{B_{s}} F_{a n n}^{L L}\left(2 a_{3}-2 a_{5}-\frac{1}{2} a_{7}+\frac{1}{2} a_{9}\right)\right]+\left[\eta_{n} \leftrightarrow \omega\right]\right\}\\
&-\frac{\sin \theta}{\sqrt{2}}\left\{  V_{u b} V_{u s}^{*}\bigg[f_{\omega} F_{B_{s} \rightarrow \eta_{s}}^{L L'}\left(a_{2}\right)+M_{B_{s} \rightarrow \eta_{s}}^{L L'}\left(C_{2}\right)\bigg]\right. \\
& - V_{t b} V_{t s}^{*}\left[f_{\omega} F_{B_{s} \rightarrow \eta_{s}}^{L L'}\left(2 a_{3}+2 a_{5}+\frac{1}{2} a_{7}+\frac{1}{2} a_{9}\right)\right. \\
& \left.\left.\left.+M_{B_{s} \rightarrow \eta_{s}}^{L L'}\left(2 C_{4}+\frac{1}{2} C_{10}\right)-M_{B_{s} \rightarrow \eta_{s}}^{S P'}\left(2 C_{6}+\frac{1}{2} C_{8}\right)\right]\right\}\right\}.
\end{aligned}
\end{equation}

\begin{equation}
\begin{aligned}
A\left(\bar{B}_{s}^{0} \rightarrow \eta'\phi\left(\phi \rightarrow K^{+} K^{-}\right) \right)=&\frac{G_{F} p_{\bar{B}_{s}^{0}} \cdot \sum_{\lambda=0, \pm 1} \epsilon(\lambda) g_{\phi} \epsilon^{*}(\lambda) \cdot\left(p_{k^+}-p_{k^-}\right)}{ \sqrt{2} s_{\phi}} \\
&\times   \left\{ \frac{\sin \theta}{\sqrt{2}}\bigg\{ V_{u b} V_{u s}^{*}\left[f_{n} F_{B_{s} \rightarrow \phi}^{L L}\left(a_{2}\right)+M_{B_{s} \rightarrow \phi}^{L L}\left(C_{2}\right)\right] \bigg.\right.\\
&- V_{t b} V_{t s}^{*}\left[f_{n} F_{B_{s} \rightarrow \phi}^{L L}\left(2 a_{3}-2 a_{5}-\frac{1}{2} a_{7}+\frac{1}{2} a_{9}\right)\right. \\
& \left.\left.+M_{B_{s} \rightarrow \phi}^{L L}\left(2 C_{4}+\frac{1}{2} C_{10}\right)+M_{B_{s} \rightarrow \phi}^{S P}\left(2 C_{6}+\frac{1}{2} C_{8}\right)\right]\right\}\\
&+\cos \theta\left\{- V_{t b} V_{t s}^{*}\left[f_{s} F_{B_{s} \rightarrow \phi}^{L L'}\left(a_{3}+a_{4}-a_{5}+\frac{1}{2} a_{7}-\frac{1}{2} a_{9}-\frac{1}{2} a_{10}\right)\right.\right. \\
& +M_{B_{s} \rightarrow \phi}^{S P'}\left(C_{6}-\frac{1}{2} C_{8}\right)+f_{B_{s}} F_{a n n}^{L L'}\left(a_{3}+a_{4}-a_{5}+\frac{1}{2} a_{7}-\frac{1}{2} a_{9}-\frac{1}{2} a_{10}\right) \\
& +M_{a n n}^{L L'}\left(C_{3}+C_{4}-\frac{1}{2} C_{9}-\frac{1}{2} C_{10}\right)-f_{B_{s}} F_{a n n}^{S P'}\left(a_{6}-\frac{1}{2} a_{8}\right) \\
& \left.\left.\left.-M_{a n n}^{L R'}\left(C_{5}-\frac{1}{2} C_{7}\right)-M_{a n n}^{S P'}\left(C_{6}-\frac{1}{2} C_{8}\right)\right]+\left[\eta_{s} \leftrightarrow \phi\right]\right\}\right\}.
\end{aligned}
\end{equation}

\begin{equation}
\begin{aligned}
A\left(\bar{B}_{s}^{0} \rightarrow \eta' \rho^{0}\left( \rho^{0} \rightarrow K^{+} K^{-}\right) \right)=&\frac{G_{F} p_{\bar{B}_{s}^{0}} \cdot \sum_{\lambda=0, \pm 1} \epsilon(\lambda) g_{ \rho} \epsilon^{*}(\lambda) \cdot\left(p_{k^+}-p_{k^-}\right)}{ \sqrt{2} s_{ \rho}} \\
&\times   \left\{\frac{\sin \theta}{2}\left\{-V_{t b} V_{t s}^{*}\left[f_{B_{s}} F_{a n n}^{L L}\left(\frac{3}{2} a_{9}-\frac{3}{2} a_{7}\right)+M_{a n n}^{L L}\left(\frac{3}{2} C_{10}\right)-M_{a n n}^{S P}\left(\frac{3}{2} C_{8}\right)\right]\right.\right. \\
& \left.+ V_{u b} V_{u s}^{*}\bigg[f_{B_{s}} F_{a n n}^{L L}\left(a_{2}\right)+M_{a n n}^{L L}\left(C_{2}\right)\bigg]+\left[\rho^{0} \leftrightarrow \eta_{n}\right]\right\}\\
&+\frac{\cos \theta}{\sqrt{2}}\left\{ V_{u b} V_{u s}^{*}\bigg[f_{\rho} F_{B_{s} \rightarrow \eta_{s}}^{L L'}\left(a_{2}\right)+M_{B_{s} \rightarrow \eta_{s}}^{L L'}\left(C_{2}\right)\bigg]\right. \\
&\left. \left.- V_{t b} V_{t s}^{*}\left[f_{\rho} F_{B_{s} \rightarrow \eta_{s}}^{L L'}\left(\frac{3}{2} a_{7}+\frac{3}{2} a_{9}\right)+M_{B_{s} \rightarrow \eta_{s}}^{L L'}\left(\frac{3}{2} C_{10}\right)-M_{B_{s} \rightarrow \eta_{s}}^{S P'}\left(\frac{3}{2} C_{8}\right)\right]\right\} \right\}.
\end{aligned}
\end{equation}

\begin{equation}
\begin{aligned}
A\left(\bar{B}_{s}^{0} \rightarrow \eta' \omega\left( \omega \rightarrow K^{+} K^{-}\right) \right)=&\frac{G_{F} p_{\bar{B}_{s}^{0}} \cdot \sum_{\lambda=0, \pm 1} \epsilon(\lambda) g_{ \omega} \epsilon^{*}(\lambda) \cdot\left(p_{k^+}-p_{k^-}\right)}{ \sqrt{2} s_{ \omega}} \\
&\times   \left\{\frac{\sin \theta}{2}\left\{  V_{u b} V_{u s}^{*}\bigg[f_{B_{s}} F_{a n n}^{L L}\left(a_{2}\right)+M_{a n n}^{L L}\left(C_{2}\right)\bigg]\right.\right. \\
& - V_{t b} V_{t s}^{*}\left[M_{a n n}^{L L}\left(2 C_{4}+\frac{1}{2} C_{10}\right)-M_{a n n}^{S P}\left(2 C_{6}+\frac{1}{2} C_{8}\right)\right. \\
+ & \left.\left.f_{B_{s}} F_{a n n}^{L L}\left(2 a_{3}-2 a_{5}-\frac{1}{2} a_{7}+\frac{1}{2} a_{9}\right)\right]+\left[\eta_{n} \leftrightarrow \omega\right]\right\}\\
&+\frac{\cos \theta}{\sqrt{2}}\left\{  V_{u b} V_{u s}^{*}\bigg[f_{\omega} F_{B_{s} \rightarrow \eta_{s}}^{L L'}\left(a_{2}\right)+M_{B_{s} \rightarrow \eta_{s}}^{L L'}\left(C_{2}\right)\bigg]\right. \\
& - V_{t b} V_{t s}^{*}\left[f_{\omega} F_{B_{s} \rightarrow \eta_{s}}^{L L'}\left(2 a_{3}+2 a_{5}+\frac{1}{2} a_{7}+\frac{1}{2} a_{9}\right)\right. \\
& \left.\left.\left.+M_{B_{s} \rightarrow \eta_{s}}^{L L'}\left(2 C_{4}+\frac{1}{2} C_{10}\right)-M_{B_{s} \rightarrow \eta_{s}}^{S P'}\left(2 C_{6}+\frac{1}{2} C_{8}\right)\right]\right\}\right\}.
\end{aligned}
\end{equation}
where the form factor involving $\eta_{s}$ is distinguished from $\eta_{n}$ by introducing a prime distinction in the upper right corner of F and M with respect to $\eta_{s}$.

\subsection{Input parameters}
The $V_{t b}$, $V_{t s}$, $V_{u b}$, $ V_{u s}$, $ V_{t d}$, and $ V_{u d}$ terms in the above equation are derived from the CKM matrix element within the framework of the Standard Model.
The CKM matrix, whose elements are determined through experimental observations, can be expressed in terms of the Wolfenstein parameters $A$,
$\rho$, $\lambda$, and $\eta$:
 $V_{t b} V_{t s}^{*}=\lambda$, $V_{u b} V_{u s}^{*}=A \lambda^{4}(\rho-i \eta)$, $V_{u b} V_{u d}^{*}=A \lambda^{3}(\rho-i \eta)(1-\frac{\lambda^{2}}{2})$, $V_{t b} V_{t d}^{*}=A \lambda^{3}(1-\rho+i \eta)$.
The most recent values for the parameters in the CKM matrix are $\lambda=0.22650\pm 0.00048$, $A=0.790_{-0.012}^{+0.017}$, $\bar{\rho}=0.141_{-0.017}^{+0.016}$, and $ \bar{\eta}=0.357\pm 0.011$. Here, we define $\bar{\rho}=\rho\left(1-\frac{\lambda^{2}}{2}\right)$ and $ \bar{\eta}=\eta\left(1-\frac{\lambda^{2}}{2}\right)$
 \cite{CKM}.
The physical quantities involved in the calculation are presented in the subsequent table :
\begin{table}[h]
{\large
\begin{center}
\caption{The remaining parameters \cite{ParticleDataGroup:2022pth, wol}(in the unit of GeV)}
\begin{tabular}{lllllllll}
\hline
\hline\\
  $m_{B_{s}}$ = $5.367 $          &\  & \qquad \, \ $m_{\eta}$ = $0.548  $ &\   &\qquad $ \enspace   f_{\phi}$ = $0.23 $                 &\    & \ \qquad $  \:f_{k}$ = $0.156 $ &\   & \qquad $\enspace \; f_{\omega}^{T}$ = $0.14 $  \\
$m_{K^0} = 0.498 $              &\   & \qquad \; $m_{\eta'}$ = $0.958  $ &\   & \quad \quad \,\,$ f_{\phi}^{T}$ = $0.22$           &\    &\qquad $ \enspace   f_{\rho}$ = $0.209 $ &\    &\, \, \qquad $f_{\omega}$ = $0.195 $ \\
\,  $  m_{\phi}$ = $1.019  $   \ &\   &\qquad \, $ m_{\pi^0}$ = $0.13498$ &\    &  \; \qquad  $f_{\pi}$ = $0.13 $                  &\   &\;\qquad  $f_{\rho}^{T}$ = $0.165   $         &\   & \qquad $ \enspace \,\; \Gamma_{\rho}$ = $0.15 $ \\
 \ \,$m_{\omega}$ = $0.782 $        &\   &  \qquad \;\ $m_{W}$ = $ 80.385 $           &\   & \qquad\;  $ f_{n}$ = $0.17      $             &\   &\,\qquad $f_{B_{s}}$ = $0.23  $         &\    & \enspace \, \qquad $ \Gamma_{\omega}$ = $8.49 \times 10^{-3} $\\
\  \,\! $m_{\rho}$ = $0.775 $       &\   &\qquad \ $m_{\pi^\pm}$ = $0.13957 $ &\      &\ \ \qquad $f_{s}$ = $0.14 $                       &\     &\ \qquad $ C_{F}$ = $4 / 3 $           &\   & \: \enspace  \qquad $ \Gamma_{\phi}$ = $4.23 \times 10^{-3} $\\
 \\
\hline
\hline
\end{tabular}
\label{tab:syst_uncert}
\end{center}}
\end{table}

\section{Analysis of data results}
\subsection{The direct CP violation from the mixing of three vector mesons}

\begin{figure}[h]
\centering
\includegraphics[height=6cm,width=9cm]{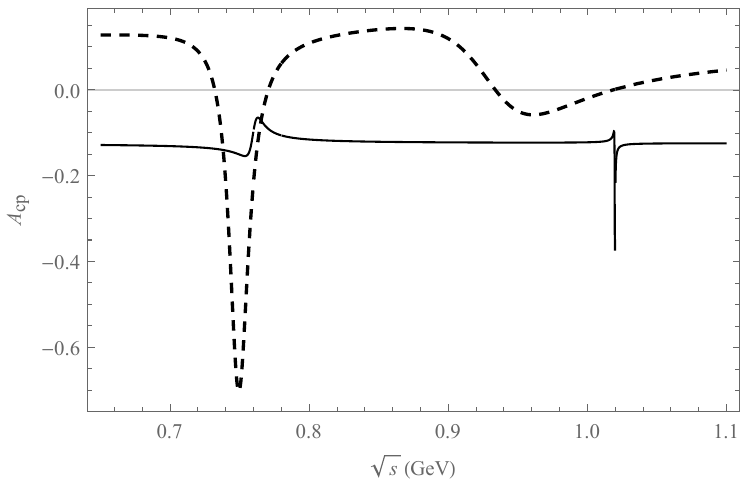}

\caption{Plot of $ A_{C P} $  as a function of  $\sqrt{s}$  corresponding to central parameter values of CKM matrix elements.  The Solid line (dashed line) corresponds to the decay channel of   $\bar B_{s} \rightarrow K^{+}K^{-} \pi (K^{0})$.}
\label{fig:2}
\end{figure}

\begin{figure}[h]
\centering

\includegraphics[height=6cm,width=9cm]{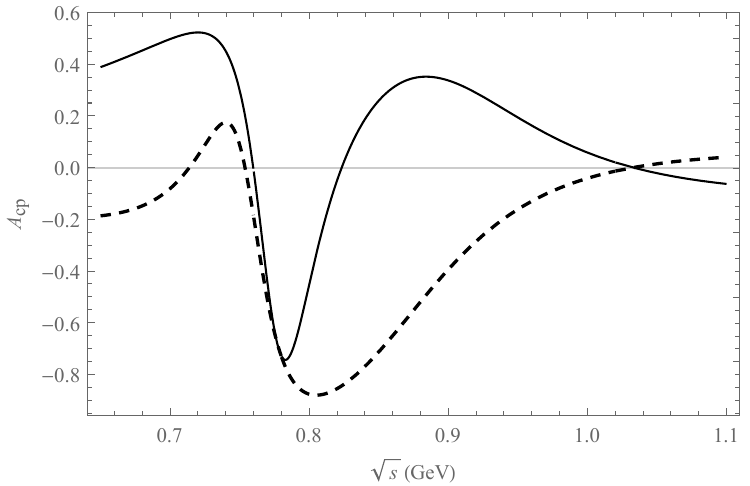}

\caption{Plot of $ A_{C P} $  as a function of  $\sqrt{s}$  corresponding to central parameter values of CKM matrix elements. The Solid line (dashed line)  corresponds to the decay channel of   $\bar B_{s} \rightarrow K^{+}K^{-} \eta (\eta' )$, respectively.}
\label{fig:3}
\end{figure}

We present the plots illustrating the CP violation in the decay processes of $\bar B_{s} \rightarrow K^{-}K^{+} P$. These plots are shown in Fig. 2 and Fig. 3, where we investigate the mixing of $\rho-\omega-\phi$ particles.
Fig. 2 and Fig. 3 depict the variation of $A_{CP}$ as a function of $\sqrt{s}$, which represents the invariant mass of $K^{+}K^{-}$. The central parameter values of CKM matrix elements are used to obtain these results. The observed $CP$ violation in these decay processes provides valuable insights into fundamental physics phenomena such as vector mesons
interferences.

The maximum of CP violation from the decay process $\bar B_{s} \rightarrow K^{+}K^{-}\pi$ in Fig.\ref{fig:2}, with a value of $-38\%$, occurs at an invariant mass of 1.02 GeV, which corresponds to the mass position of the $\phi$ meson. Additionally, small peaks are also observed in the invariant mass range of $\rho^{0}-\omega$. Therefore, it can be concluded that the decay process $\bar B_{s} \rightarrow \phi \pi \rightarrow K^{+}K^{-}\pi$ plays a significant role in this decay channel.
Furthermore, for the decay process $\bar B_{s} \rightarrow  K^{+}K^{-} K^{0}$, a sharp variation in CP violation is observed when the invariant masses of $K^{+}K^{-}$ pairs fall within the region around 0.75 GeV, reaching a peak value of $-70\%$. In this case, it is effects from the $\rho^{0}-\omega$ mixing mechanism rather than contributions from the QCD penguin dominant decay  $\bar B_{s} \rightarrow \phi K^0$. Consequently, interference effects are expected to occur within a range near 0.7 GeV- 0.8 GeV.  It should be noted that only the tree graph contributes to the $\bar B_{s} \rightarrow  \phi K^{0}$ decay. However, the mixed resonance effect between $\phi-\omega-\rho$ leads to a smaller violation peak shift in the invariant mass position of the $\phi$ meson.

While the decay process $\bar B_{s} \rightarrow  K^{+}K^{-}\eta (\eta')$ is more intricate, we first consider the decay process $\bar B_{s} \rightarrow  V \eta(\eta')$ involving $\eta(\eta')$. The physical states of $\eta$ and $\eta'$ mesons are composed of a mixture of flavor eigenstates, namely $\eta_{n}$ and $\eta_{s}$. Furthermore, there is no contribution from penguin graphs in the decay process $\bar B_{s} \rightarrow  \phi \eta_{s}$; hence, the amplitude contribution of the decay $B_s\to K^+K^-\eta (\eta^\prime)$ within this entire mixture is negligible. As depicted in Fig.\ref{fig:3}, resonant interplay between large CP violation in both invariant mass intervals ($\rho^{0}-\omega $ and $\phi $) ultimately leads to the observed effect shown in Fig.\ref{fig:3}.
In the figure, it is evident that the CP violation peak in $\bar B_{s} \rightarrow  K^{+}K^{-} \eta(\eta')$  occurs with a magnitude of $-74\%$ ($-88\%$) near the range $0.8$ GeV. This observation allows us to comprehend the trend of CP violation in these decay processes, which is advantageous for our research. Additionally, we can determine the invariant mass value of the $K^{+}K^{-}$ pair during significant CP violation events, providing an opportunity for experimental measurement.

\subsection{Numerical results of the localized integrated CP asymmetry}

The relationship between CP violation and invariant mass in the decay process, as derived from the preceding section, provides valuable insights into the dynamics of CP violation. However, to comprehensively investigate regional CP violation and establish for future experiments, we perform a local integration analysis of CP violation within the studied decay process. Consequently, Table II presents the localized CP violation for the aforementioned decay processes.

\begin{table}[h]
{\renewcommand
\scalebox{12}
\centering %
\renewcommand{\arraystretch}{2} %
\setlength{\tabcolsep}{4mm}{
\begin{center}
\caption{The peak local$(0.98GeV\leq \sqrt{s}\leq1.06GeV)$ integral of  $\mathrm{A}^{\Omega} _{\mathrm{CP}}$ from different resonance rangs for $\bar B_{s} \rightarrow  K^{+} K^{-} \pi $ $(K^{0},\eta,\eta')$ decay processes.}
\begin{tabular}{ ccccc  }
\hline
Decay channel                                     & $\phi-\rho-\omega$ mixing             &$\phi-\rho$ mixing                     & $\phi-\omega$ mixing               & $\rho-\omega$ mixing \\ \hline
$\bar B_{s} \rightarrow K^{+} K^{-} \pi^{0}$       & $\mathrm{-0.124\pm 0.012}$            & $\mathrm{-0.126\pm 0.008} $             &$\mathrm{-0.147\pm0.004}  $            &$ \mathrm{-0.124\pm 0.010}$ \\
 $\bar B_{s} \rightarrow K^{+} K^{-} K^{0}$       & $\mathrm{-0.001\pm 0.000} $          &$\mathrm{0.0003\pm 0.0001} $               &$\mathrm{0.0008\pm0.0002} $          &$ \mathrm{0.169\pm 0.004}$    \\
$\bar B_{s} \rightarrow K^{+} K^{-} \eta$       & $\mathrm{0.021\pm0.0001}  $          & $\mathrm{0.0174\pm 0.0002}$                & $\mathrm{0.010\pm0.001}$         & $\mathrm{-0.237\pm 0.007}$    \\
$\bar B_{s} \rightarrow K^{+} K^{-} \eta'$     & $ \mathrm{-0.014\pm 0.005}$           & $\mathrm{-0.012\pm 0.008}$              & $\mathrm{-0.007\pm 0.002}$       &$\mathrm{-0.240\pm 0.005}$    \\
\hline
\end{tabular}
\end{center}}}
\end{table}

According to Table II, the integration range (0.98 GeV-1.06 GeV) corresponds to the threshold of $V \rightarrow K^{+}K^{-}$ decay process. The resonance effect between different particles can lead to more pronounced CP violation phenomena in various energy intervals. However, considering the threshold effect for generating $K^{+}K^{-}$ meson pairs, we provide the local integral values as shown in Table II. To compare the similarities and differences between three-particle and two-particle resonance effects, we also present the local integral results of CP value under two-particle resonance in Table II.

In the $\bar B_{s} \rightarrow K^{+} K^{-} \pi^{0}$ decay process, the value of CP violation changes less in the resonance regions above the threshold values due to any two-particle or three-particle mixing. Although the mixed resonance contributes a peak value of $-38\%$ for $\bar{B}_{s}^{0} \rightarrow K^{+} K^{-} \pi^{0}$ decay process in Fig. 2, the local integral values have minimal variation within a specific range in comparison to the overall resonance interval.
The values of $A_{CP}^{\Omega}$ exhibit a consistent magnitude of approximately 0.124.

The values of $A_{CP}^{\Omega}$ are small due to the contributions from $\phi-\rho-\omega$ mixing, $\phi-\rho$ mixing, and $\phi-\omega$ mixing. However, a significant CP violation of 0.169 can be observed from the contribution of $\rho-\omega$ mixing. This behavior changes in the decay process $\bar B_{s} \rightarrow K^{+} K^{-} K^{0}$ since it involves the QCD penguin dominant decay $\bar B_{s} \rightarrow \phi K^{0}$ without any tree-level contribution. In this case, only the decay process involving intermediate states with $\rho -\omega$ particles exhibits noticeable CP violation.

The decay process $\bar B_{s} \rightarrow K^{+} K^{-} \eta(\eta')$ is also a special decay process, characterized by the presence of meson mixing between $\eta$ and $\eta'$.
The process $\bar B_{s} \rightarrow \phi \eta_s$ is the QCD penguin dominant decays without any contribution from a tree diagram, while the process $\bar B_{s} \rightarrow \phi \eta_n$ involves  tree diagram and penguin diagram contributed.
Thus $\eta_{s}$ and $\eta_{n}$ mixing results in the presence of a smaller tree contribution for $\eta$ ($\eta'$).
Consequently, the involvement of $\phi$ as an intermediate state in the decay process leads to a reduction in the value of $A_{CP}^{\Omega}$.
The CP violation induced by the decay process involving $\rho-\omega$ mixing exhibits distinct characteristics, with a maximum value of $-0.237 (-0.240)$ observed for the processes $\bar B_{s} \rightarrow K^{+} K^{-} \eta$ ($\bar B_{s} \rightarrow K^{+} K^{-} \eta'$), respectively.

Theoretical errors give rise to uncertainties in the results. In general, the major theoretical uncertainties arise from power corrections beyond the heavy quark limit, necessitating the inclusion of $1/m_b$ power corrections. Unfortunately, there exist numerous possible $1/m_b$ power suppressed effects that are typically nonperturbative in nature and therefore not calculable using perturbation theory. Consequently, this scheme introduces additional sources of uncertainty. The first error arises from variations in the CKM parameters, while the stems from hadronic parameters, such as the shape parameters, form factors, decay constants, and the wave function of the Bs meson.   The third error corresponds to selecting appropriate hard scales that characterize the size of next-to-leading order QCD contributions. By employing central values for these parameters, we initially compute numerical results for CP violation and subsequently incorporate errors based on standard deviation in Table II.
It has been determined that the impact of mixing parameter errors on local CP violation is negligible compared to the overall CP asymmetry, therefore this influence value will not be further discussed.

\section{Summary and conclusion}

The CP violation in the decay process of $\bar{B}_{s}^{0}$ meson is predicted through an invariant mass analysis of $K^{+} K^{-}$ meson pairs within the resonance region, resulting from the mixing of $\phi$, $\omega$, and $\rho$ mesons. We observe a sharp change in CP violation within the resonance regions of these mesons. Local CP violation is quantified by integrating over phase space. For the decay process $\bar B_{s} \rightarrow K^{+} K^{-} \pi^{0}$, we find a local CP violation value around $-0.12$ arising from interference between $\phi$, $\omega$, and $\rho$ mesons. In decays such as  $\bar B_{s} \rightarrow K^{+} K^{-} K^{0}$, $\bar B_{s} \rightarrow K^{+} K^{-} \eta$ and $\bar B_{s} \rightarrow K^{+} K^{-} \eta'$, CP violations are observed due to contributions from both two-meson mixing and three-meson mixing processes. Particularly involving the $\rho$ - $\omega$ mixing, the local CP violation is large.
Experimental detection of local CP violation can be achieved by reconstructing the resonant states of $\phi$, $\omega$, and $\rho$ mesons
within the resonance regions.

 We propose a quasi-two-body approach, namely, $\bar{B}_{s}^{0}\rightarrow VP \rightarrow K^{+} K^{-} P$ to elucidate the three-body decay mechanism of $\bar{B}_{s}^{0}\rightarrow K^{+} K^{-} P$.
During this process, V acts as an intermediate state and undergoes resonance with other particles, ultimately decaying into pairs of $K^{+} K^{-}$ mesons.
The three-body decay process of bottom is appropriately formulated using the chain decay of quasi-two-body. We consider the $B\rightarrow RP_3$ decay process as a case study for analyzing quasi-two-body decays, where R represents an intermediate resonance state that can further decay into harons $P_{1,2}$, while $P_3$ denotes another final hadron.
The process under consideration can be factorized utilizing the narrow width approximation (NWA).
The expression for $B\rightarrow RP_3$ can be written as follows: $\mathcal{B} \left( B\rightarrow RP_3\rightarrow P_1P_2P_3 \right) =\mathcal{B} \left( B\rightarrow RP_3 \right) \mathcal{B} \left( B\rightarrow P_1P2 \right)$ which holds true due to the branching ratio.
The effects of small widths $\phi$, $\rho$, and $\omega$ in quasi-two-body decay processes into $KK$ can be safely neglected.
Considering the substantial decay rate of $\rho(770)$, it is reasonable to perform a correction.
From the QCD factorization approach, the correction factor for the decays process of $B^{-}\rightarrow \rho(770)\pi^{-}\rightarrow \pi^{+}\pi^{-}\pi^{-}$ is at level $7\%$.
The parameter $\eta _R$ is introduced as a quant approximation between $\varGamma \left( B\rightarrow RP_3 \right) \mathcal{B} \left( B\rightarrow P_1P_2 \right)$ and $\varGamma \left( B\rightarrow RP_3\rightarrow P_1P_2P_3 \right)$ \cite{chenghaiyang2021prd,chenghaiyang2021plb}.
When calculating the CP violation, this constant can be eliminated, thereby exerting no influence on our ultimate outcome.

Recently, the LHCb experimental group has made significant progress in investigating the three-body decay of B mesons and has obtained noteworthy findings \cite{LHCb:2022fpg}. By analyzing previous experimental data, they have measured direct CP violation in various decay modes such as $B^{\pm} \rightarrow K^{+} K^{-} K^\pm$, $B^{\pm} \rightarrow \pi^{+} \pi^{-} K^\pm$, $B^{\pm} \rightarrow \pi^{+} \pi^{-} \pi^\pm$, and $B^{\pm} \rightarrow K^{+} K^{-}\pi^\pm$.  Based on LHCb experiments, it is anticipated that future investigations will primarily focus on exploring the three-body decays of $\bar{B}_s$.

 \section*{Acknowledgements}
 This work was supported by  Natural Science Foundation of Henan (Project No. 232300420115) .


\end{spacing}

\begin{thebibliography}{99}
\bibitem{Cabibbo:1963yz}N. Cabibbo, Phys. Rev. Lett. {\bf 10}, 531(1963).

\bibitem{LHCb:2012kja}R. Aaij, et al. [LHCb Collaboration]. LHCb-CONF-2012-018(2012).
\bibitem{LHCb:2012uja}R. Aaij, et al. [LHCb Collaboration]. LHCb-CONF-2012-028(2012).
\bibitem{Wang:2015ula}C. Wang, et al. Eur. Phys. J. C {\bf 75} no.11, 536(2015).
\bibitem{LHCb:2013ptu} R. Aaij, et al. [LHCb Collaboration]. Phys. Rev. Lett. {\bf 111}, 101801 (2013).

\bibitem{Ali:2007ff} Ahmed Aliet, et al. Phys. Rev. D {\bf 76}, 074018(2007).
\bibitem{xiao2007xc} Zhen-Jun Xiao, Dong-qin Guo, and Xin-fen Chen, Phys. Rev. D {\bf 75}, 014018(2007) .
\bibitem{Hua:2020usv}J. Hua, et al. Phys. Rev. D {\bf 104},016025(2021).
\bibitem{Zou:2020fax}Z. T. Zou, et al. Eur. Phys. J. C {\bf 80}, 517(2020).
\bibitem{BaBar:2014zli}J. P. Lees, et al. [BaBar]. Phys. Rev. Lett. {\bf 113} no.20, 201801(2014).
\bibitem{Bertacchi:2023jzv}V.Bertacchi. [Belle-II]. Nucl. Part. Phys. Proc. {\bf 324-329 }, 107-112(2023).
\bibitem{CLEO:2007vpk}N. E. Adam, et al. [CLEO]. Phys. Rev. Lett. {\bf 99},041802(2007) .
\bibitem{Aaijprl2013} R. Aaij, et al. [LHCb Collaboration]. Phys. Rev. Lett.  {\bf 111}, 101801 (2013).
\bibitem{PMplb1981} P.M. IVANOV, et al. Phys. Lett. B {\bf 107}, 297 (1981).
\bibitem{Achasov2016} M. N. Achasov, et al,  Phys. Rev. D {\bf 94}, 112006 (2016).

\bibitem{Lu:2022rdi}Gang L\"{u}, et al. Chin. Phys. C {\bf46}, 113101(2022).
\bibitem{Lv2023epj} Gang L\"{u}, et al. Eur. Phys. J. C {\bf83}, 345 (2023).


\bibitem{CE2009} C. E. Wolfe, K. Maltman, Phys. Rev. {\bf D80}, 114024 (2009).
\bibitem{CE2011} C. E. Wolfe, K. Maltman, Phys. Rev. {\bf D83}, 077301 (2011).
\bibitem{Lu:2018fqe}Gang L\"{u}, et al. Phys. Rev. D {\bf 98}, no.1, 013004 (2018).
\bibitem{MN2000} M. N. Achasov, et.al., Nucl. Phys. {\bf B569}, 158 (2000).


\bibitem{Chen:1999nxa} Y. H. Chen, et al. Phys. Rev. D {\bf60}, 094014(1999).

\bibitem{Wolfe:2010gf}  C.E.Wolfe, et al. Phys. Rev. D {\bf83}, 077301(2011).
\bibitem{Wolfe:2009ts}  C.E.Wolfe, et al. Phys. Rev. D {\bf80}, 114024(2009).
\bibitem{Bruch:2004py}C.Bruch, et al. Eur. Phys. J. C {\bf39}, 41-54 (2005).


\bibitem{Guo:2000uc} X. H. Guo, et al. Phys. Rev. D {\bf63}, 056012 (2001).

\bibitem{Zhang:2013oqa} Z.H.Zhang, et al. Phys. Rev. D {\bf87}, no.7, 076007 (2013).



\bibitem{Li:2006jv} Hsiang-nan Li, et al. Phys. Rev. D  {\bf 74}, 094020(2006).

\bibitem{CKM}L. Wolfenstein, Phys. Rev. Lett. {\bf51},1945 (1983).
\bibitem{ParticleDataGroup:2022pth}R. L. Workman et al. (Particle Data Group). Review of Particle Physics, PTEP 2022, 083C01(2022).
\bibitem{wol}L. Wolfenstein, Phys. Rev. Lett. {\bf13}, 562 (1964).







\bibitem{chenghaiyang2021prd} Hai-Yang Cheng, Cheng-Wei Chiang, Chun-Khiang Chua, Phys. Rev. {\bf D103}, 036017 (2021).
\bibitem{chenghaiyang2021plb} Hai-Yang Cheng, Cheng-Wei Chiang, Chun-Khiang Chua, Phys. Lett. {\bf B813}, 136058 (2021).
\bibitem{LHCb:2022fpg} R. Aaij, et al.[LHCb]. Phys. Rev. D {\bf108}, no.1, 012008 (2023).
























\end{thebibliography}
\end{document}